\documentclass{emulateapj}

\usepackage{aas_macros}
\usepackage{natbib}

\newcommand{\hkpc}{$h^{-1}{\ }{\rm kpc}$}
\newcommand{\hMpc}{$h^{-1}{\ }{\rm Mpc}$}
\newcommand{\hMsun}{$h^{-1}{\ }{\rm M_{\odot}}$}

\newcommand{\sigmabox}{\sigma_B}

\newcommand{\Table}[1]{Table~\ref{#1}}
\newcommand{\Sec}[1]{Section~\ref{#1}}

\newcommand{\Fig}[1]{Fig.~\ref{#1}}

\begin{document}

\title{On the starting redshift for cosmological simulations: Focusing on halo properties}

\author{Alexander Knebe\altaffilmark{1,2}, 
Christian Wagner\altaffilmark{2},
Steffen Knollmann\altaffilmark{1,2},
Tobias Diekershoff\altaffilmark{2},
Fabian Krause\altaffilmark{2}
}

\altaffiltext{1}{Departamento de Fisica Teorica,
Modulo C-XI,
Facultad de Ciencias,
Universidad Autonoma de Madrid,
28049 Cantoblanco, Madrid,
Spain}

\altaffiltext{2}{Astrophysikalisches Institut Potsdam,
				An der Sternwarte 16,
				14482 Potsdam,
				Germany}

\begin{abstract}
  We systematically study the effects of varying the starting redshift
  $z_i$ for cosmological simulations \textit{in the highly non-linear
    regime}. Our primary focus lies with the (individual) properties
  of dark matter halos -- namely the mass, spin, triaxiality, and
  concentration -- where we find that even substantial variations in
  $z_i$ leave only a small imprint, at least for the probed mass range
  $M\in[10^{10}, 10^{13}]$\hMsun\ and when investigated at redshift
  $z=0$.  We further compare simulations started by using the standard
  Zel'dovich approximation to runs based upon initial conditions
  produced with second order Lagrangian perturbation theory. Here we
  observe the same phenomenon, i.e. that differences in the studied
  (internal) properties of dark matter haloes are practically
  undetectable. These findings are (for the probed mass range) in
  agreement with other work in the literature. We therefore conclude
  that the commonly used technique for setting up cosmological
  simulations leads to stable results at redshift $z=0$ for the mass,
  the spin parameter, the triaxiality, and the concentration of dark
  matter haloes.
\end{abstract}


\section{Introduction} \label{sec:introduction}
During the past two decades numerical simulations of cosmic structure
formation have become a standard tool in cosmology. Advances in
computational algorithms combined with an ever increasing size of the
machines used to run these simulations have made it possible to
simulate regions of the universe with unprecedented dynamic range,
recently culminating in multi-billion particle simulations of galactic
halos \citep[e.g.,][]{Stadel08, Springel08}. However, most of the
efforts towards achieving this goal have been related to refining the
simulation techniques and studying the differences between various
codes, respectively \citep[cf.][]{Frenk99, Knebe00, OShea05, Regan07,
  Heitmann07, Agertz07, Tasker08}. But common to all these simulations
-- irrespective of the applied code -- is the way how the initial
conditions (ICs) are generated: practically everyone in the field uses
the so-called Zel'dovich approximation \citep[][ZA]{Zeldovich70} first
employed for cosmological simulation by \citet{Klypin83} and
\citet{Efstathiou85}, despite the notion that other methods may be
more suitable and accurate, respectively \citep[e.g.,][]{Pen97,
 Sirko05, Hansen07, Joyce08}. We like to remind the reader that the
ZA is a first order Lagrangian perturbation theory and hence we will
also refer to ZA as ``lpt1''. There are only a few parameters that
require specification for this method, one of which is the starting
redshift $z_i$; for an elaborate discussion of (most of) the
parameters influencing the ICs we refer the reader to \citet{Joyce08}.

There is already a great deal of work out there related to the
credibility of the ZA with respects to transients from initial
conditions \citep[e.g.][]{Scoccimarro98, Crocce06b, Tatekawa07}, the
memory of initial conditions \citep{Crocce06a}, errors in real-space
statistical properties of the ZA \citep[e.g.][]{Knebe03, Sirko05},
discreteness effects \citep{Baertschiger01, Joyce07a, Joyce07b} as
well as descriptions of how to accurately generate multi-mass
simulations \citep[e.g.][]{Navarro97, Prunet08}. However, to our
surprise the literature yet lacks a systematic study of the right
choice for the starting redshift $z_i$ with respects to the internal
properties of dark matter halos at today's time.

On the one hand, workers in the field argue that the validity of the
linear theory -- upon which the ZA is based -- is enforced by choosing
a starting redshift in such a way that the resulting variance of the
discrete density field (from now on referred to as $\sigmabox$, see
definition below) is significantly less than unity
\citep[e.g.,][]{Crocce06b, Prunet08}. But how much smaller than unity
exactly? As a rule of thumb people adopted a value of
$\sigmabox\approx0.1-0.2$ \citep[as pointed out by, for instance,][]{Crocce06b,
  Lukic07, Prunet08}.\footnote{Please note that these references do
  not argue in favour of $\sigmabox<<1$; they rather refer to this
  choice as ``common practice''.} But what is this choice based upon?

On the other hand, \citet{Lukic07}, for instance, provide a useful
formula for estimating the starting redshift $z_i$ based upon the
requirement that the fundamental mode in the simulation box is well in
the linear regime \citep[Eq.(20) in ][]{Lukic07}; and they
additionally discuss the necessity to further restrict the initial
displacement of particles to a level that the first crossing of
trajectories happens several expansion factors after starting the
simulation \citep[see also ][]{Valageas02, Crocce06b}. They argue that
it is important to allow for a sufficient number of expansion factors
between the starting redshift and the highest redshift of physical
significance to ensure that artifacts from the initial conditions
(e.g. a regular grid structure) are lost. This goes along wih the
tests presented in \citet{Reed03} where it is argued that the
simulation should be evolved for at least an expansion factor
of~$\sim10$ before extracting (mass function) measurements. Starting
too late (or allowing for too few expansion before extracting physical
information from the simulation) will delay collapse of the first
halos acting as seeds for further structure formation, e.g. the
Zeldovich approximation cannot account for shell-crossing wherein mass
piles up as it flows towards overdensities \citep[cf., ][]{Jenkins01,
 Reed03, Heitmann08}. But what is ``too late''?

However, one should also not start too early to avoid numerical
round-off errors and shot noise of the particles used to sample the
primordial matter density field (e.g.~in the case of glass ICs)
\citep[e.g., ][]{Lukic07}. The main objective of this paper is to
shed light on this issue and study the differences in properties of
dark matter halos at redshift $z=0$ when systematically varying the
starting redshift $z_i$ and consequently $\sigmabox$.

We though have to acknowledge that this subject has in part been
touched upon by other people \citep[e.g.,][]{Jenkins01, Reed03,
  Heitmann06, Crocce06b, Lukic07, Tinker08, Joyce08}. However, the
primary focus of these studies was solely the mass (function) of dark
matter halos and numerical effects in generating the ICs,
respectively. In that regards, it has been shown by, for instance,
\citet{Reed03} that the starting redshift can have a substantial
influence on the high-redshift mass function. However, such effects
should have evolved away by lower redshifts since the tiny fraction of
matter that is in halos at high redshift is soon incorporated into
clusters or large groups \citep{Reed03}. This is actually confirmed by
the findings of \citet{Jenkins01} who observed that the mass function
at low-$z$ is not very sensitive to the starting redshift. A similar
result was found by \citet{Tinker08} for halos with mass
$M<10^{14}$\hMsun: they compared mass functions obtained with
simulations based upon ZA initial conditions but variations in the
starting redshift from $z_i=60$ to $z_i=35$. However, the difference
increased for halos more massive than $10^{14}$\hMsun\ and could
become as large as 10-20\%, nevertheless depending on the code and
particulars of the simulation, respectively (cf. their Fig.14). This
goes along with the results of \citet{Crocce06b} who found a
$\sim10$\% discrepancy at $10^{15}{\rm  M_{\odot}}$ in $z=0$ mass
functions when comparing the standard first-order ZA with second-order
Lagrangian perturbation theory (from now on referred to as ``lpt2'')
for generating initial conditions for cosmogical simulations. This may
be caused by the fact that ZA assumes straight lines for particle
trajectories whereas lpt2 includes the effects of gravitational tides;
and the latter are most pronounced for regions containing the rarest
peaks of largest height that tend to evolve into the largest galaxy
clusters at low redshift \citep{Tinker08}.

Here we extend and complement all previously mentioned studies by
quantifying the impact of the initial starting redshift $z_i$ upon the
\textit{individual properties of gravitationally bound objects} at
redshift $z=0$ other than the mass alone. In addition, the same
comparison is done for simulations started by a 2nd order Lagrangian
perturbation method (lpt2). This complements the research carried out
by \citet{Crocce06b} who presented an in-depth investigation of
transients from initial conditions in cosmological simulations in ZA
and lpt2. In contrast to that study, we consider scales that are much
deeper in the non-linear regime.

\section{The Simulations} \label{sec:simulations}

We ran a series of simulations with the publicly available
\texttt{GADGET2} code. All simulations consist of $N=256^3$ particles
in a cubical volume of 25\hMpc\ side length. The cosmology we imposed
is compliant with the latest WMAP results \citep[i.e. $\Omega_0=0.28,
\Omega_\Lambda=0.72, h=0.73, \sigma_8=0.76, n_s=0.96$;
][]{Komatsu08}. The force resolution of our simulations is 2\hkpc. As
the phases in the generation of the ICs were identical across all
models we are able to cross-compare both individual particles as well
as individual halos. The latter were identified with the MPI-enabled
open source halo finder \texttt{AHF}\footnote{\texttt{AHF} is already
  freely available from \url{http://www.aip.de/People/aknebe}}
(\texttt{AMIGA}'s-Halo-Finder, Knollmann \& Knebe 2009), which is
based upon the \texttt{MHF} halo finder of \citet{Gill04a}: halos are
located as peaks in an adaptively smoothed density field of the
simulation using an adaptive grid hierarchy based upon a refinement
criterion that matches the force resolution of the actual simulation
(i.e. in our case 5 particles per cell); local potential minima are
computed for each of these peaks and the set of particles that are
gravitationally bound to the halo are returned. For every halo we
calculate a suite of canonical properties (e.g. velocity, mass, spin,
shape, concentration, etc.)  based upon the particles within the
virial radius. The virial radius $R_{\rm vir}$ is defined as the point
where the density profile (measured in terms of the cosmological
background density $\rho_b$) drops below the virial overdensity
$\Delta_{\rm vir}$, i.e.  $M(<R_{\rm vir})/(4\pi R_{\rm vir}^3/3) =
\Delta_{\rm vir} \rho_b$. This threshold $\Delta_{\rm vir}$ is based
upon the dissipationless spherical top-hat collapse model and is a
function of both cosmological model and time. For the given cosmology
it amounts to $\Delta_{\rm vir}=354$ at $z=0$.

We need to mention that we restricted our analysis to objects with at
least 100 particles corresponding to a lower mass cut of $M_{\rm
  min}=7.2\times10^{9}$\hMsun\ and according to the study presented in
\citet{Knollmann09} we are certain that our halo catalogue is complete
at this level. The total number of objects found is of order 20000
with approximately 10 objects with $M\geq 10^{13}$\hMsun. For a more
elaborate discussion of how well our numerically determined mass
functions agree with existing fitting formulae in the literature we
refer the interested reader to the actual \texttt{AHF} paper by
\citet{Knollmann09}.

The simulations can be split into two distinct sets classifying the
method applied to displace the particles from a regular lattice at the
initial redshift $z_i$, i.e. 
using the code by \citet{Sirko05} 
we either apply 

\begin{itemize}
\item lpt1: the Zel'dovich approximation or
\item lpt2: 2nd order Lagrangian perturbation theory.
\end{itemize}

In each of these sets we varied the starting redshift and used the
values $z_i=150, 100, 50$, and $25$ and our convention for labelling
these runs is summarized in \Table{tab:labels}. That table further
lists the starting redshift alongside the aforementioned rms variance
of the matter distribution inside the computational volume

\begin{equation} \label{eq:SigmaBox}
\sigmabox^2 = \frac{1}{2\pi^2} \int_{k_{\rm min}}^{k_{\rm max}} P(k) k^2 dk
\end{equation}

\noindent
where $k_{\rm min} = 2\pi/B$ represents the fundamental mode
determined by the box size $B$ and $k_{\rm max}=\pi N^{1/3}/B$ the
Nyquist frequency that additionally depends on the number of particles
$N$ used for the initial conditions.

\begin{deluxetable}{ccc}
\tablecaption{Simulation labels alongside initial conditions parameters.
\label{tab:labels}} 
\tablewidth{0pt} 
\tablehead{ \colhead{run} & \colhead{$z_i$} & \colhead{$\sigma_{\rm box}$} }
\startdata
150-lpt1 & 150 & 0.05 \\
100-lpt1 & 100 & 0.07 \\
050-lpt1 & 50 & 0.14 \\
025-lpt1 & 25 & 0.28 \\
150-lpt2 & 150 & 0.05 \\
100-lpt2 & 100 & 0.07 \\
050-lpt2 & 50 & 0.14 \\
025-lpt2 & 25 & 0.28 \\
\enddata
\end{deluxetable}

\section{Density Field Comparison} \label{sec:densityfield}
Even though the primary focus of this study lies with the internal
properties of dark matter halos (to be presented in \Sec{sec:halos}
below) we start with an investigation of the effects of the starting
redshift (and order of the Lagrangian scheme) upon the matter density
field.  This is due to the lack of differences found in
\Sec{sec:halos} and we hence considered it mandatory to verify our
simulations against the results obtained by other workers in the field
who predominantly explored the power spectrum (and mass function of
halos) in that regards.

According to the commonly accepted and applied method to generate ICs
for cosmological simulations we refer to model ``lptX-050'' as our
reference model for which the rms of the matter field $\sigmabox=0.14$
is in the commonly used range of $0.1-0.2$. We therefore compare all
other models in the first set (``lpt1'' set) against this particular
run and in analogy use ``lpt2-050'' as the reference in the lpt2 set.

\subsection{Applied Comparisons} \label{sec:comparisons}
All models in a given lpt1/2 set are compared against the reference
model started at $z=50$. To gauge the influence of using first
(lpt1/ZA) or second order (lpt2) Lagrangian perturbation theory in the
generation of the ICs we additionally cross-compare runs of these two
sets against each other that started at the same redshift. In summary,
this leaves us with ten comparisons summarized in
\Table{tab:comparisons}.

\begin{deluxetable}{ccc}
\tablecaption{Applied Corss-Comparisons.
\label{tab:comparisons}} 
\tablewidth{0pt} 
\tablehead{ \colhead{comparison} & \colhead{model \#1} & \colhead{model \#2} }
\startdata
\#1  & 150-lpt1 & 050-lpt1 \\
\#2  & 100-lpt1 & 050-lpt1 \\
\#3  & 025-lpt1 & 050-lpt1 \\
\#4  & 150-lpt2 & 050-lpt2 \\
\#5  & 100-lpt2 & 050-lpt2 \\
\#6  & 025-lpt2 & 050-lpt2 \\
\#7  & 150-lpt1 & 150-lpt2 \\
\#8  & 100-lpt1 & 100-lpt2 \\
\#9  & 050-lpt1 & 050-lpt2 \\
\#10 & 025-lpt1 & 025-lpt2 \\
\enddata
\end{deluxetable}

\subsection{Power Spectra} \label{sec:powerspectra}
\begin{figure}
\includegraphics[angle=-90,width=0.45\textwidth]{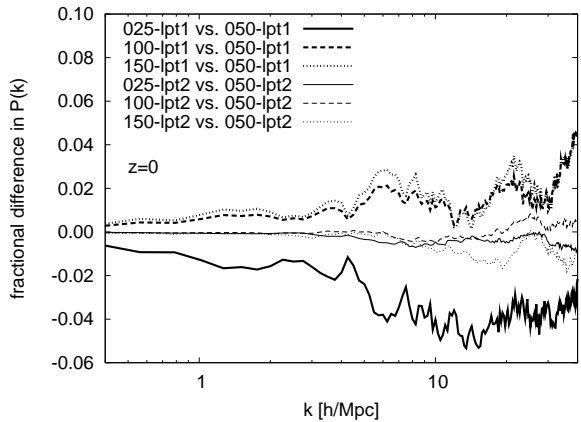}
\caption{Fractional difference in the power spectrum P(k) 
regarding the different starting redshifts 
	at redshift $z=0$. The wavenumber $k$ ranges from the largest mode to the Nyquist frequency of the particle grid.}
\label{fig:ps_start}
\end{figure}

\begin{figure}
\includegraphics[angle=-90,width=0.45\textwidth]{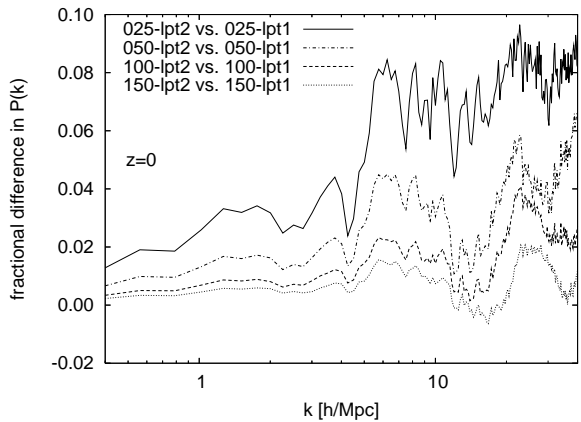}
\caption{Fractional difference in the power spectrum P(k) regarding the different schemes (ZA and lpt2) used to set up the initial conditions	at redshift $z=0$.}
\label{fig:ps_lpt2}
\end{figure}

We start our comparison with the power spectrum
of matter density fluctuations. The power spectrum 

\begin{equation}
P(k) \propto \left\langle |\delta(\vec k)|^2\right\rangle \ ,
\end{equation}

\noindent
where $ \delta(\vec k)$ is the Fourier transform of the density
contrast $\delta(\vec x)$, i.e.

\begin{equation}
 \delta(\vec k) = \int {\frac{d^3x}{(2\pi)^3} \exp (i \vec k \cdot \vec x) \delta(\vec x)} \ ,
\end{equation}

\noindent
is the commonly used statistic to describe the clustering of the
density field.  Here, we compute it by applying an FFT on a regular
$1024^3$ grid using the cloud-in-cell \citep[][CIC]{Hockney88} scheme
for the mass assignment.

We compare the power spectra of the different runs by calculating the
fractional difference, e.g. in the case of ``150-lpt1 vs. 050-lpt1''
the fractional difference in $P(k)$ is given by $(P(k)_{\rm
150-lpt1}-P(k)_{\rm 050-lpt1})/P(k)_{\rm 050-lpt1}$.

The results for the runs started at different initial redshifts $z_i$
with respect to the reference run started at $z_i=50$ (i.e.,
comparisons \#1 through \#6 in \Table{tab:comparisons}) are shown in
\Fig{fig:ps_start}. The thick and thin lines correspond to the ZA and
lpt2 runs, respectively. We notice that the runs which were set up
using the ZA differ more strongly ($\pm 4\%$) when changing the
respective starting redshift than the runs set up with lpt2 ($\pm
1\%$). Or in other words, ICs generated using the ZA are more
sensitive to the actual starting redshift (at least with respects to
the power spectrum analysis) than ICs based upon lpt2. We further
find, that for the ZA runs, the earlier the simulation started the
more power we get at $z=0$ compared to the reference model started at
$z=50$, especially at the small-scale/high-$k$ end. Both of these
results are compliant with the findings of \citet{Crocce06b} and
\citet{Ma07}. However, we like to note that given our box size and
particle number we probe much smaller scales and hence a clustering
regime that is highly non-linear.

In \Fig{fig:ps_lpt2} we compare the power spectra of the lpt2 runs
against the power spectra of the lpt1 runs for each starting redshift
$z_i$ at redshift $z=0$ (i.e.~comparisons \#7 through \#10 of
\Table{tab:comparisons}). We find that, in general, the lpt2 initial
conditions lead to more power than the ZA initial
conditions. Obviously, the effect is bigger the later the simulation
started ($\sim 7\%$ for $z_i=25$ and $\sim 1\%$ for $z_i=150$), simply
because the difference between first order and second order Lagrangian
perturbation theory decreases with redshift. And again, this
observation agrees with the results of other works that compared lpt1
against lpt2 \citep[e.g.][]{Crocce06b, Ma07, Nishimichi08, Heitmann08}.

The bottomline of this sub-section therefore is that -- whatever we
will find in the subsequent study below -- our simulations reproduce
the same trends as found by others when varying either the starting
redshift or the order of the Lagrangian perturbation theory, at least
when it comes to studying the (power spectrum of the) matter density
field.

\subsection{Particle Positions} \label{sec:particlepositions}
\begin{figure}
\includegraphics[width=0.45\textwidth]{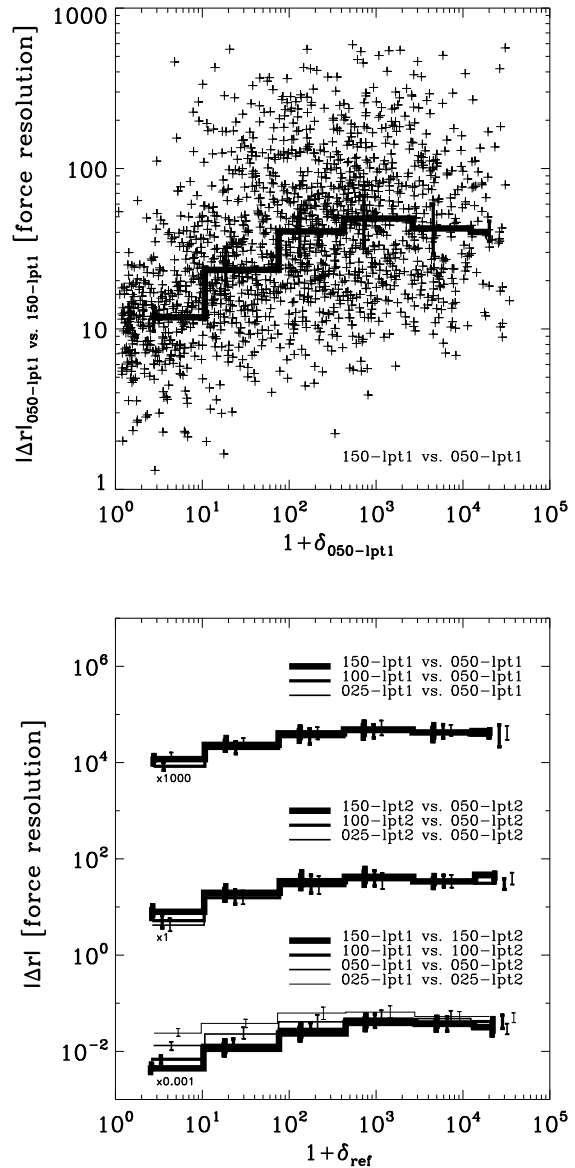}
\caption{Deviation (normalized to the force resolution) of particle
 coordinates at $z=0$. The upper panel presents a random sample of
 1\% of all particles alongside the median $\Delta r$ in six bins in
 $1+\delta_{\rm ref}$. The lower panel only shows the medians
 multiplied by 100, 1, and 0.01, respectively, to avoid crowding.}
\label{fig:deltar}
\end{figure}

\begin{deluxetable*}{cccccc}
\tablecaption{Cross-correlation coefficients.
\label{tab:ccc}} 
\tablewidth{0pt} 
\tablehead{ \colhead{comparison} & \colhead{$L=32^3$} & \colhead{$L=64^3$} & \colhead{$L=128^3$} & \colhead{$L=256^3$} & \colhead{$L=512^3$}}
\startdata
150-lpt1 vs. 050-lpt1 & 0.9853 & 0.9452 & 0.8755 & 0.7487 & 0.6800 \\
100-lpt1 vs. 050-lpt1 & 0.9867 & 0.9487 & 0.8839 & 0.7599 & 0.6882 \\
025-lpt1 vs. 050-lpt1 & 0.9851 & 0.9232 & 0.8758 & 0.7477 & 0.6637 \\
\\
150-lpt2 vs. 050-lpt2 & 0.9861 & 0.9424 & 0.8690 & 0.7566 & 0.6833 \\
100-lpt2 vs. 050-lpt2 & 0.9872 & 0.9461 & 0.8749 & 0.7650 & 0.6911 \\
025-lpt2 vs. 050-lpt2 & 0.9874 & 0.9462 & 0.8758 & 0.7662 & 0.6932 \\
\\
150-lpt1 vs. 150-lpt2 & 0.9889 & 0.9523 & 0.8856 & 0.7732 & 0.6982 \\
100-lpt1 vs. 100-lpt2 & 0.9879 & 0.9503 & 0.8808 & 0.7659 & 0.6914 \\
050-lpt1 vs. 050-lpt2 & 0.9851 & 0.9518 & 0.8673 & 0.7447 & 0.6793 \\
025-lpt1 vs. 025-lpt2 & 0.9808 & 0.8824 & 0.8484 & 0.7175 & 0.6380 \\
\enddata
\end{deluxetable*}

As all ICs were generated using the same phases we are in the
advantageous position to compare individual particle positions across
models. To this extent we apply two tests. The first consists of
calculating the modulus of the difference between those positions

\begin{equation}
|\Delta r| = |\vec{r}_i - \vec{r}_j|
\end{equation}

\noindent
where $\vec{r}_i$ is the position in simulation $i$ and $\vec{r}_j$ in
simulation $j$, and the second utilizes the so-called density
cross-correlation coefficient.

In \Fig{fig:deltar} we show for our set of comparisons
(cf. \Sec{sec:comparisons}) the position difference $|\Delta r|$ (in
units of the force resolution, i.e.~2\hkpc) for 1\% of particles randomly selected
from the total number of particles as a function of density as
measured at the position of the particle\footnote{The density contrast
 $\delta = (\rho-\overline{\rho})/\overline{\rho}$ has first been
 calculated by assigning the mass of each particle to a regular grid
 of size $512^3$. Then the grid values have been interpolated back to
 the particles' positions.} in the reference model started at
redshift $z_i=50$ (i.e.~050-lpt1 and 050-lpt2, respectively) and hence
labelled $\delta_{\rm 050-lpt1}$ in the upper panel and $\delta_{\rm
 ref}$ in the lower panel.  The upper panel shows the actual scatter
plot for one particular comparison (i.e., 050-lpt1 vs. 150-lpt1) together with the median
while the lower panel shows only the medians of $\Delta r$ in six logarithmically
spaced bins.  Note that the whole particle set has been used to
calculate the medians and to avoid crowding in the figures we
multiplied the medians by $100$, $1$, and $0.01$, respectively. The
error bars represent the $25^{\rm th}$ and $75^{\rm th}$ percentiles 
(slightly shifted for each model along the x-axis for clarity).

We notice the expected trend for $|\Delta r|$ to increase with
increasing density contrast, i.e.~the differences across models are
more pronounced in high-density regions. We now checked (though not
shown here) that the differences are never larger than the virial
radii of the halos these particles reside in. However, the observed
trend is expected: the origin of these deviations is the dynamical
instability of particle trajectories in the high-density regions
\citep[e.g.][]{Knebe00, Valluri07}. As is well known, the trajectories
within virialized systems tend to be chaotic and any small differences
existing at any time moment will tend to grow very fast with time. The
divergence can thus be expected to be more important in non-linear
regions and this explains the trend of larger $\Delta r$'s in denser regions. 
The differences in the low-density regions are substantially
smaller, but still larger than the the force resolution and hence
considered physical. However, when investigating underdense regions
$1+\delta<1$ we notice that the median of $|\Delta r|$ ``saturates''
at approximately the level of $10\times$force resolution and hence
defines the level that marks the minimum expectation for the position
difference.

We also observe that the medians do \textit{not} show considerable
variations when changing the starting redshift. Further, the trend for
$|\Delta r|$ to increase with $\delta$ -- that appears to be
independent of $z_i$ -- is also of comparable strength for lpt1 and
lpt2.

However, when comparing lpt1 against lpt2 there appears to be a drift
towards smaller particle position differences (in low-density regions)
when moving to higher starting redshifts. This is readily explained by
the fact that at higher redshifts the differences between lpt1 and
lpt2 vanish. Nevertheless, this trend is far less pronounced in
high-density regions.
We conclude that the differences in $P(k)$ as seen in
\Fig{fig:ps_lpt2} therefore stem from rather low-density regions.

We also cross-compared an earlier started model to a later started one
at its actual starting redshift, e.g. a snapshot of run 150-lpt1 at
redshift $z=50$ to the ICs of simulation 050-lpt1. We though chose to
not show the results as all differences in the positions $|\Delta r|$
are smaller than the force resolution and we hence consider them
unphysical.

We like to caution the reader that this particular test of
investigating the spatial differences $|\Delta r|$ does not provide us
with the information which method is better or worse. It simply shows
that the differences in the final matter distribution are practically
all at the same level, irrespective of starting redshift and order of
the Lagrangian perturbation theory used. There is a trend for
differences in the particles' position to increase in high-density
regions but this trend is the same for cross-comparisons of the lpt1,
lpt2 and lpt1 vs. lpt2 models, respectively. 

As a second quantitative measure of differences we construct the
so-called cross-correlation coefficient \citep{Coles93, Splinter98,
 Knebe00}

\begin{equation}
K = \frac{\langle\delta_i \delta_j\rangle}{\sigma_i \sigma_j} \ ,
\end{equation}
where $i$ and $j$ specify the different simulations and the 
average is taken over the computational box. 

To compute $K$, we have calculated the densities $\delta_i, \delta_j$
on a regular mesh using the triangular-shaped cloud
\citep[][TSC]{Hockney88} density assignment scheme, and then used the
resulting density field to compute $\langle\delta_i \delta_j\rangle$
as well as the corresponding variances $\sigma_i$ and $\sigma_j$. We
have varied the size of the grid in order to show the dependence of
the cross-correlation on the smoothing scale of the density field.

The results are shown in \Table{tab:ccc} where it is obvious that in
all cases the cross-correlation worsens for smaller smoothing scales
(larger grid sizes). This is compliant with the trends seen in
\Fig{fig:deltar}, i.e.~that $|\Delta r|$ increases with $\delta$: the
smaller the smoothing, the smaller structures are resolved in the
density field. The degraded cross-correlation therefore indicates that
there are differences in locations and/or densities of these
small-scale structures. 
If we restrict the correlation analysis to a coarse grid, we smooth the particle
distribution with a fairly large smoothing length and smear out the
details and differences in the small-scale structure.

But nevertheless, the absolute values of $K$ are of the same order
when contrasting different comparisons. One
may additionally argue that $K$ is slightly higher for the
lpt2~vs.~lpt2 cross-correlations than for the corresponding
lpt1~vs.~lpt1 case indicative of the higher order of the lpt2 scheme, 
which is in agreement with our findings for the power spectra (cf. \Fig{fig:ps_start}).
Further -- as in the case of
the power spectra (cf. \Fig{fig:ps_lpt2}) -- the cross-correlations
for lpt1~vs.~lpt2 become smaller for earlier starting redshifts, e.g. applying a $256^3$ grid the $K$ value for 150-lpt1~vs.~150-lpt2 is 8\% higher than for 025-lpt1~vs.~025-lpt2. This is exactly the same value as we find for the difference of the two power spectra at the corresponding wave number (i.e. the Nyquist frequency).

We also like to note that the value of the density cross-correlation
coefficient $K$ found here is of the same order as for a comparison of
\textit{the same} simulation run with different codes under matching
conditions \citep[cf.\ Table~2 in][]{Knebe00}.

In summary, we do observe differences in the particle positions when
changing either the starting redshift or the order of the Lagrangian
perturbation theory. And these differences are obviously more
pronounced in high-density regions where particle trajectories appear
to be more ``chaotic''. However, trends for these differences to
change with starting redshift are at best marginal, irrespective of
lpt1 or lpt2.

\section{Halo Comparison} \label{sec:halos}
\begin{figure}
\includegraphics[width=0.45\textwidth]{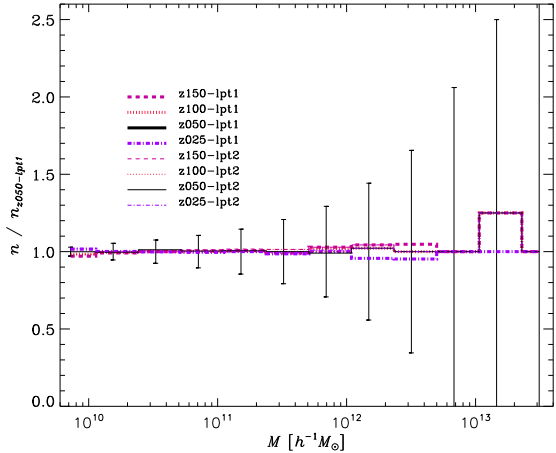}
\caption{Ratio of the mass function at redshift $z=0$ of gravitationally bound objects for all eight models to the respective reference model started at $z=50$. The (Poissonian) error bars measure the $3-\sigma$ variance.
\label{fig:massfuncRatio}}
\end{figure}

\begin{figure}
\includegraphics[width=0.45\textwidth]{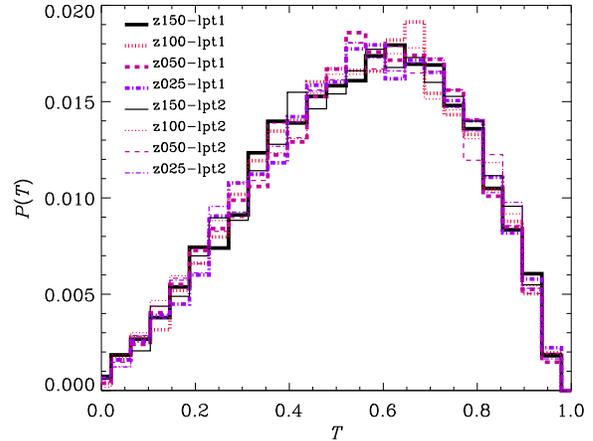}
\includegraphics[width=0.45\textwidth]{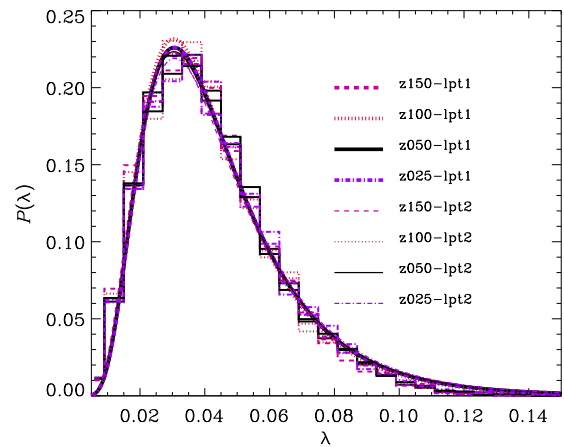}
\includegraphics[width=0.45\textwidth]{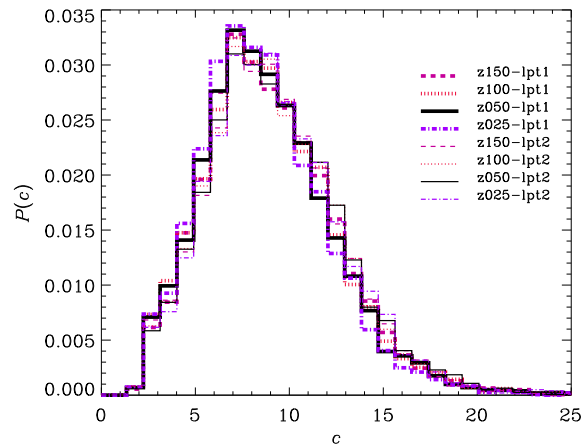}
\caption{Probability distribution of the triaxiality parameter $T$
 (top), the spin parameter $\lambda$ (middle), and the concentration
 $c$ (bottom) for all models.}
\label{fig:PTspinE}
\end{figure}

The same model comparisons as in the previous section
(cf. \Table{tab:comparisons}) are now going to be performed using the
halo catalogues obtained via the halo finder \texttt{AHF}
(cf. \Sec{sec:simulations}). In that regards we will focus on four
quantities, namely the mass $M$, the triaxiality \citep{Franx91}

\begin{equation}
T = \frac{a^2-b^2}{a^2-c^2}
\end{equation}

\noindent
where $a>b>c$ are the eigenvalues of the moment of inertia tensor,
the spin parameter \citep{Bullock01b}

\begin{equation}
\lambda = \frac{L}{\sqrt{2} \ M V R}
\end{equation}

\noindent
where $L$ is the absolute value of the angular momentum, $M$ the halo
mass, $R$ its radius, and $V^2=GM/R$, and finally the concentration

\begin{equation}
c = \frac{R}{R_2}
\end{equation}

\noindent
where $R_2$ measures the point where $r^2 \rho(r)$ peaks, with
$\rho(r)$ representing the spherically averaged density profile.

\subsection{Distributions} \label{sec:distributions}
\begin{deluxetable}{ccc}
\tablecaption{Best-fit parameters for the spin parameter distributions.
\label{tab:Pspin}} 
\tablewidth{0pt} 
\tablehead{ \colhead{run} & \colhead{$\lambda_0$} & \colhead{$\sigma_0$} }
\startdata
150-lpt1 & 0.040 & 0.509 \\
100-lpt1 & 0.039 & 0.501 \\
050-lpt1 & 0.040 & 0.507 \\
025-lpt1 & 0.040 & 0.506 \\
150-lpt2 & 0.039 & 0.524 \\
100-lpt2 & 0.040 & 0.515 \\
050-lpt2 & 0.040 & 0.513 \\
025-lpt2 & 0.040 & 0.519 \\
\enddata
\end{deluxetable}

We start with inspecting the distribution functions for the four
quantities under investigation. For the mass function $n(>M)$ we
decided to plot the ratios with respects to the reference model
started at $z_i=50$ whereas for $P(T)$, $P(\lambda)$, and $P(c)$ we
chose to plot the actual distributions. The results can be viewed in
Figs.~\ref{fig:massfuncRatio} and~\ref{fig:PTspinE}.

The dependence of the mass function on the starting redshift has
already been investigated by several groups before \citep[e.g.,
][]{Jenkins01, Reed03, Crocce06b, Lukic07, Tinker08}.

Our own results can be viewed in \Fig{fig:massfuncRatio} where we plot
the ratio of the mass function of a particular model and the fiducial
050-lpt1/2 run; the (Poissonian) error bars measure the 3-$\sigma$
variance and are hence proportional to $3\times\sqrt{N_{\rm
    halos}^{\rm bin}}$ where $N_{\rm halos}^{\rm bin}$ is the average
number of halos in the respective bin. 

The (lack of) differences seen in \Fig{fig:massfuncRatio} is
consistent with the outcome of similar studies in the field
\citep[e.g.][]{Jenkins01, Reed03, Crocce06b, Lukic07, Tinker08}: most
of the previous investigations focussed on the very high-mass end of
the mass function, i.e. $10^{14}-10^{15}$\hMsun\ and found transients
to be important there. However, that mass regime is not probed by our
simulations; we are analysing halos within a mass range for which none
of the previously mentioned papers have found any differences either.

The limited influence of the starting redshift upon (low-mass) dark
matter halos at redshift $z=0$ can also be observed for the shape,
spin parameter, and concentration presented in \Fig{fig:PTspinE}:
there are hardly any noticable differences in the distributions when
changing $z_i$. This figure is accompanied by \Table{tab:Pspin} for
which the spin parameter distributions have been fitted to a lognormal
function

\begin{equation} \label{lognormal}
P(\lambda) = \displaystyle \frac{1}{\lambda \sqrt{2\pi\sigma_0^2}}
\exp \left( {-\frac{\ln^2 (\lambda/\lambda_0)}{2 \sigma_0^2}} \right) \ ,
\end{equation}

\noindent
with the two best-fit parameters $\lambda_0$ and $\sigma_0$ listed in
\Table{tab:Pspin}. We again note that they are practically
indistinguishable, irrespective of the model and the starting redshift
$z_i$. However, we note that the width of the distribution as measured
by $\sigma_0$ appears to be marginally smaller in the lpt1 set.

Our results from this section indicate that the starting redshift has
practically no influence on today's attributes of dark matter halos --
at least not for the properties analysed here, namely the mass, the
spin parameter, the triaxiality, and the concentration and for objects
within the given mass range $10^{10}-10^{13}$\hMsun. And the same
holds for the order of the Lagrangian perturbation theory,
i.e.~whether lpt1 or lpt2 is used to generate the ICs has no effect on
the particulars of halos at redshift $z=0$.

\subsection{Cross-Correlations} \label{sec:crosscorrelations}
\begin{figure}
\includegraphics[width=0.45\textwidth]{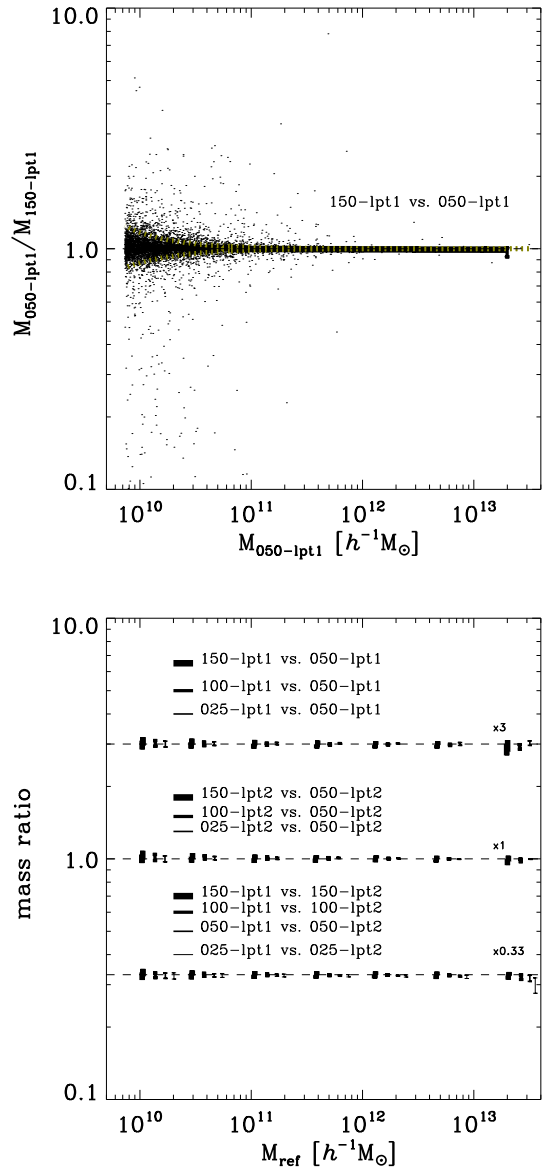}
\caption{Cross-correlation of the mass $M_{\rm vir}$ of individual
 halos.  The upper panel presents a random sample of 1\% of all
 particles alongside the median mass ratio in six bins in $M_{\rm
   ref}$. We further show the curves for mass ratio stemming
 from a differences in the number of particles of $\pm 20$ as dashed
 curves. The lower panel only shows the $25^{\rm th}$ and the $75^{\rm th}$ percentiles as bars around the medians multiplied by 3,
 1, and 1/3, respectively, to avoid crowding.
 \label{fig:MassMass}}
\end{figure}

\begin{figure}
\includegraphics[width=0.45\textwidth]{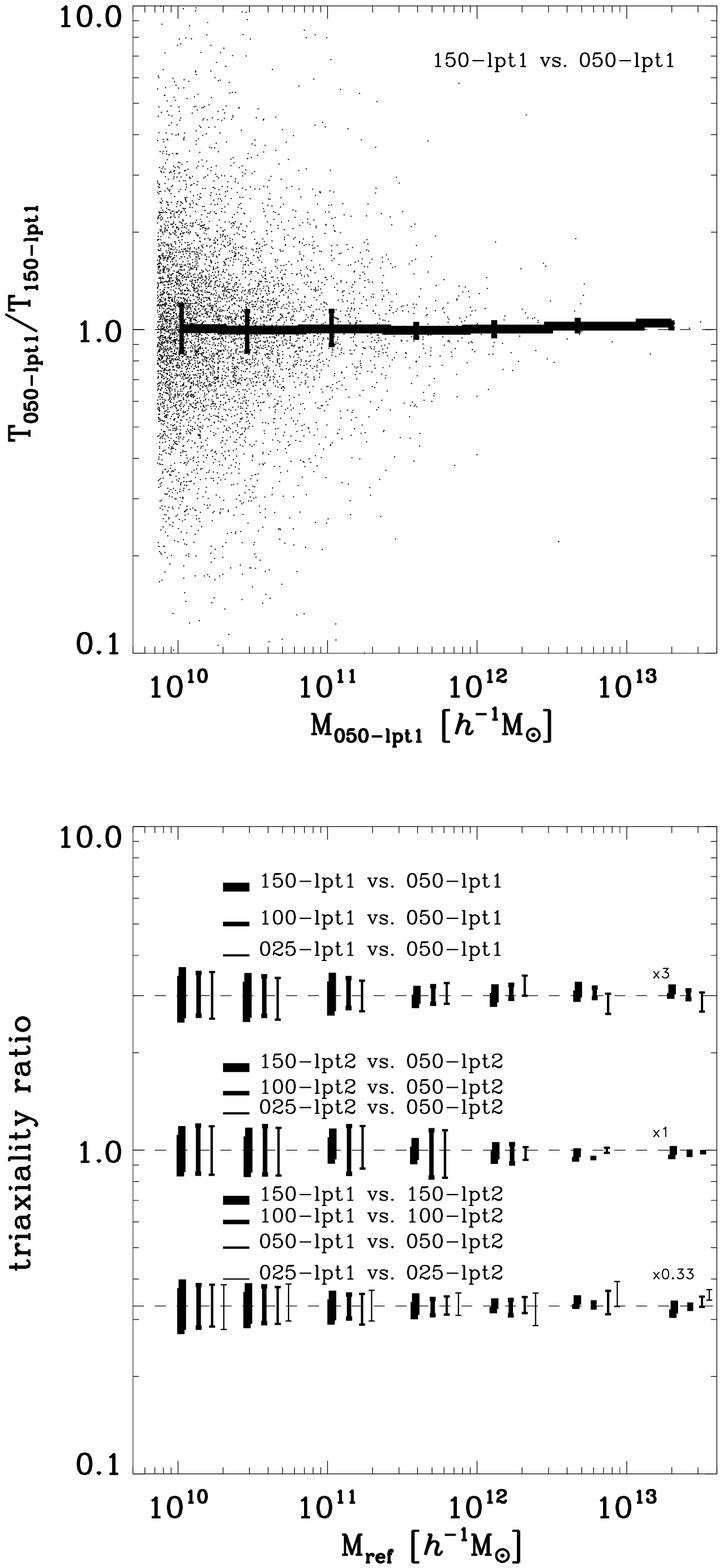}
\caption{Cross-correlation of the triaxiality parameter $T$ of individual halos. 
The same logic for the panels is used as in \Fig{fig:MassMass}.
\label{fig:TT}}
\end{figure}

\begin{figure}
\includegraphics[width=0.45\textwidth]{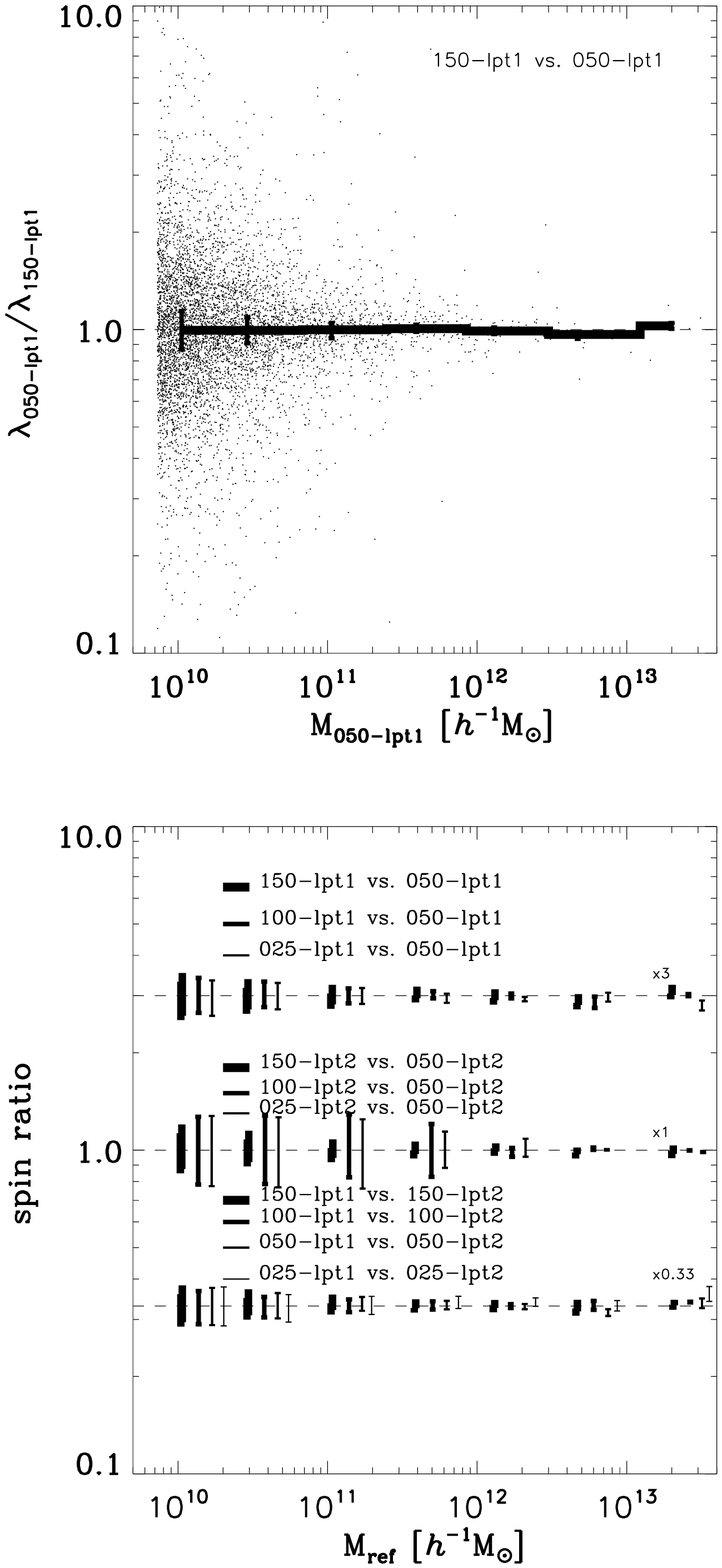}
\caption{Cross-correlation of the spin parameter $\lambda$ of individual halos.
The same logic for the panels is used as in \Fig{fig:MassMass}.
\label{fig:LamLam}}
\end{figure}

\begin{figure}
\includegraphics[width=0.45\textwidth]{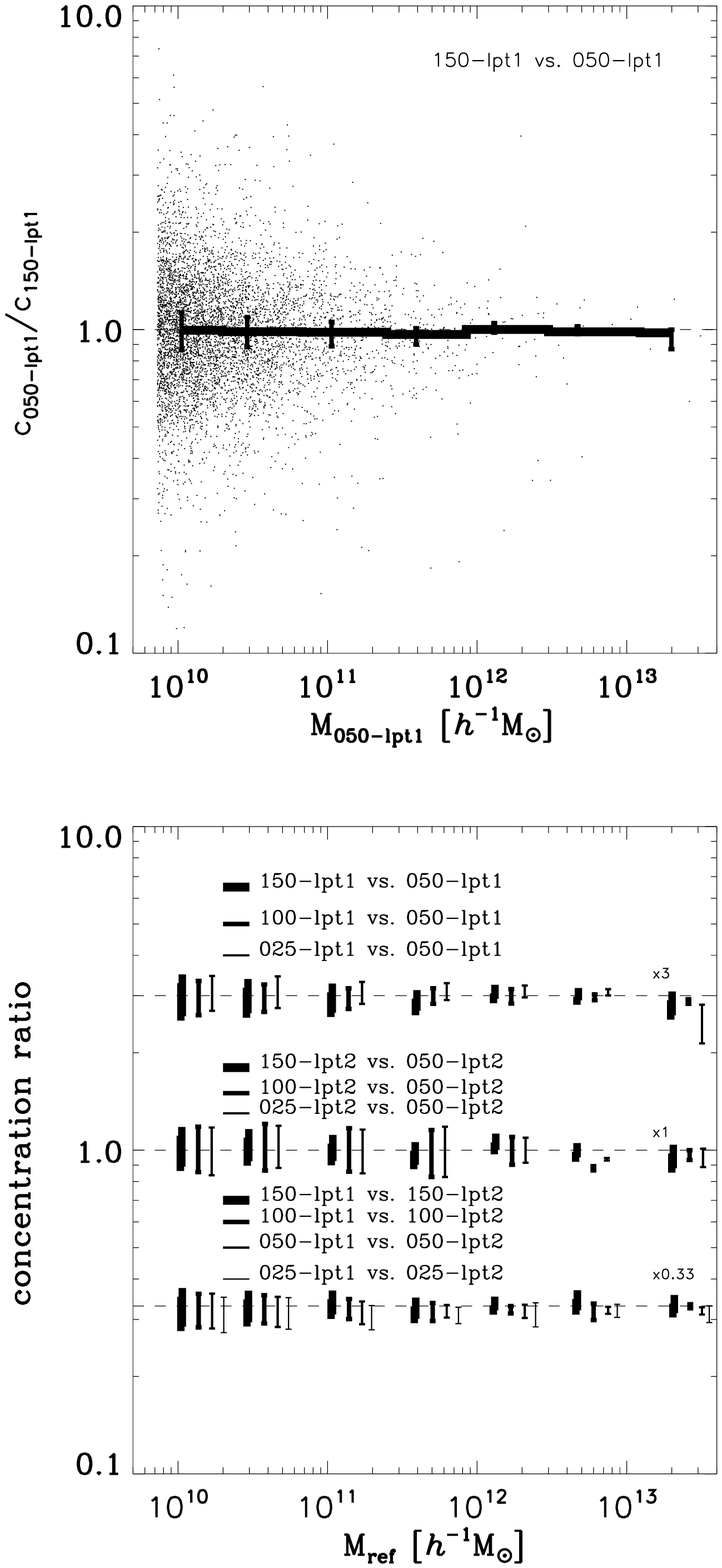}
\caption{Cross-correlation of the concentration $c$ of individual halos.
The same logic for the panels is used as in \Fig{fig:MassMass}.
\label{fig:c5c5}}
\end{figure}

As the simulations were started with identical phases we could use the
particle IDs to establish a mapping between two different
simulations. This has been applied in \Sec{sec:particlepositions}
where we presented a direct comparison of individual particles (e.g.,
the spatial difference $|\Delta r|$).  But if we plan to do the same
for the halos, we require a more sophisticated technique to uniquely
cross-identify halos amongst different simulations. To this extent we
utilize a tool that comes with the \texttt{AHF} package and is called
\texttt{MergerTree}. Originally it serves the purpose of identifying
corresponding objects in the same simulation at different redshifts
(and hence the name \texttt{MergerTree}). But it can also be applied
to simulations of different models run with the same initial phases
for the initial conditions like in our case. The \texttt{MergerTree}
cross-correlation is done by linking objects that share the most
common particles and has been succesfully applied to similar
comparisons before \citep[e.g.][]{Knebe06happi}.

We consider again mass $M$, triaxiality $T$, spin parameter $\lambda$,
and concentration $c$ and the results can be found in
Figs.~\ref{fig:MassMass}--\ref{fig:c5c5} where we plot the ratio of
said quantities to the reference model against the mass $M$ in that
reference model. As in \Fig{fig:deltar} we present in the upper panel
the actual scatter plot for 1\% of the particles alongside the median
in seven bins while the lower panel this time only shows the range from the 
$25^{\rm th}$ to the $75^{\rm th}$ percentile 
(centered about the median multiplied by 3, 1, and 1/3,
respectively) marginally shifted for each model on the $x$-axis for
clarity.

We note that (the medians of) the ratios in all instances are
consistent with unity. The ``error'' bars -- representing 50\% of the halos -- are well concentrated around unity, in particular in the case of the mass ratio, and do not vary strongly with halo mass.
We though observe that the individual scatter about that median (cf. upper panels) increases with decreasing halo mass;
;as a matter of fact, the
ratio can become substantially large (of order two for the mass and
order ten for the shape/spin) for the lowest mass objects. However, 
we attribute this to the differences in the particle positions (in
high-density regions and hence in and about halos) already noted in
\Fig{fig:deltar}. And as shown in \citet{Knollmann09}, tiny variations
in the particle positions lead to different density contours upon
which the halo finding algorithm of \texttt{AHF} is based. This then
entails marginal differences in the halo properties that become more
apparent at the low mass end where halos only consist of few
particles. To better gauge this explanation we also plot in
\Fig{fig:MassMass} as dashed lines those curves that mark a difference
in particle number by 20; and we observe that most of the ratios lie
between these curves.

The result of this section is rather remarkable as the ratio of
properties of cross-identified halos is \textit{always} consistent
with unity, even if the starting redshift is as small as $z_i=25$. The
observed scatter about unity increases at the lower mass end which is
naturally explained by variations in the number of particles making up
the actual halo and hence is rather a peculiarity of the halo finding
algorithm \citep{Knollmann09}. However, the variation of the median as
indicated by the error bars in the lower panels of
Figs.~\ref{fig:MassMass}--\ref{fig:c5c5} clearly shows that the
medians are consistent with unity.

We are therefore confident that while there are one-to-one variations
the statistical properties of halos are unaffected by the starting
redshift (cf. \Sec{sec:distributions}). This result is again extended
to the order of the Lagrangian perturbation theory used for generating
the ICs. While we still found remnants from transients in the power
spectrum (cf. \Fig{fig:ps_lpt2}) and deviations of particle positions
(cf. \Fig{fig:deltar}) no effects are observed for individual
properties of dark matter halos at redshift $z=0$ anymore.

\section{Conclusions and Discussion} \label{sec:conclusions}
We performed a systematic study of varying the starting redshift $z_i$
for cosmological simulations. We further used two methods to generate
the ICs, namely the Zel'dovich approximation (a first order Lagrangian
perturbation method) and an explicit second order Lagrangian
perturbation code. The resulting snapshots at redshift $z=0$ were
analysed with respects to properties of the matter density field as
well as the statistical and individual properties of dark matter
halos. Besides of (expected) fluctuations in high-density regions and
at the low-mass end of the halo population the differences are rather
marginal. On average, the objects have indistinguishable properties
irrespective of the starting redshift. Surprisingly, even the
simulation started as late as $z_i=25$ for which the rms matter
fluctuations are $\sigmabox=0.28$ and hence larger than the commonly
adopted value of $\leq0.1-0.2$ gave comparable results to the fiducial
model started at $z_i=50$. For the probed mass range
$M\in[10^{10},10^{13}]$\hMsun, we conclude that (at least at low
redshift $z\approx 0$) the starting redshift has little (if any)
influence on the (statistical) properties of the halo population. Or
in other words, the resulting halo properties of mass $M <
10^{13}$\hMsun\ at redshift $z=0$ are stable against reasonable
variations in the rms matter fluctuations $\sigmabox$ and hence the
starting redshift $z_i$. We note that our results are in agreement
with the findings presented throughout the literature where
differences have only been found on scales $M>10^{14}$\hMsun\
\citep[e.g.,][]{Jenkins01, Reed03, Heitmann06, Crocce06b, Lukic07,
  Tinker08, Joyce08}.


We further found that also the order of the Lagrangian perturbation
theory used to generate the ICs is of little relevance. While it
leaves an imprint on the matter distribution (especially the power
spectrum) the study of gravitationally bound objects at redshift $z=0$
is hardly affected, at least when it comes to the spin parameter
$\lambda$, the triaxiality $T$, and the concentration $c$.  This
result actually goes along with the arguments presented by
\citet{Reed03}, i.e. the effects of a ``wrong'' starting redshift
should have evolved away by lower redshifts since the tiny fraction of
matter that is in halos at high redshift is soon incorporated into
clusters or large groups. This explains why \citet{Jenkins01} as well
as \citet{Tinker08} did find that the (low-$z$) mass function is not
very sensitive to the starting redshift.

One concern that may be raised with respects to our results is the
size of our simulation box and finite volume effects, respectively. We
concede that the amplitude of fluctuations at the size of the box is
of order unity at redshift $z=0$ and therefore couplings to modes on
even larger scales are missing. While this certainly affects the halo
mass function and the power spectrum we though argue that our primary
results are not influenced. If we were to adjust our mass functions
for such finite volume effects using the recipes outlined in, for
instance, \citet{Reed07}, \citet{Lukic07}, or \citet{Power06} we would
need to apply the same correction to all our models. But as we only
cross-compare models amongst each other our findings should not be
contaminated by such a systematic change. In addition, we showed in a
previous study that the influence of large-scale modes upon the
properties of dark matter haloes at redshift $z=0$ will lead to
differences in, for instance, the spin parameter of order $<15$\% with
the concentration being hardly affected at all \citep{Power06}.

We further acknowledge that our analysis focused on simulations
primarily analysed at redshift $z=0$. And our results are not as
surprisingly as one may initially think as at this time the scales
under consideration are deeply in the non-linear regime and the memory
of the initial conditions should have been lost \citep{Crocce06a}. At
earlier times the situation may be different which explains the
results of others who found a dependence of the high-mass of the
multiplicity function especially at high redshift
\citep[e.g.][]{Reed03, Crocce06b, Lukic07, Tinker08}. We leave a more
in-depth investigation of this (using higher resolution simulations)
to a future study.

\acknowledgments
AK is supported by the MICINN through the Ramon y Cajal programme.
AK and SRK acknowledge funding through the Emmy Noether programme of
the DFG (KN 755/1). The simulations and the analysis presented in this
paper were carried out on the Sanssouci cluster at the
Astrophysikalisches Institut Potsdam.


\bibliographystyle{apj}
\bibliography{archive}

\begin{thebibliography}{47}
\expandafter\ifx\csname natexlab\endcsname\relax\def\natexlab#1{#1}\fi

\bibitem[{{Agertz} {et~al.}(2007){Agertz}, {Moore}, {Stadel}, {Potter},
  {Miniati}, {Read}, {Mayer}, {Gawryszczak}, {Kravtsov}, {Nordlund}, {Pearce},
  {Quilis}, {Rudd}, {Springel}, {Stone}, {Tasker}, {Teyssier}, {Wadsley}, \&
  {Walder}}]{Agertz07}
{Agertz}, O., {Moore}, B., {Stadel}, J., {Potter}, D., {Miniati}, F., {Read},
  J., {Mayer}, L., {Gawryszczak}, A., {Kravtsov}, A., {Nordlund}, {\AA}.,
  {Pearce}, F., {Quilis}, V., {Rudd}, D., {Springel}, V., {Stone}, J.,
  {Tasker}, E., {Teyssier}, R., {Wadsley}, J., \& {Walder}, R. 2007, \mnras,
  380, 963

\bibitem[{{Baertschiger} \& {Sylos Labini}(2001)}]{Baertschiger01}
{Baertschiger}, T., \& {Sylos Labini}, F. 2001, ArXiv Astrophysics e-prints

\bibitem[{{Bullock} {et~al.}(2001){Bullock}, {Dekel}, {Kolatt}, {Kravtsov},
  {Klypin}, {Porciani}, \& {Primack}}]{Bullock01b}
{Bullock}, J.~S., {Dekel}, A., {Kolatt}, T.~S., {Kravtsov}, A.~V., {Klypin},
  A.~A., {Porciani}, C., \& {Primack}, J.~R. 2001, \apj, 555, 240

\bibitem[{{Coles} {et~al.}(1993){Coles}, {Melott}, \& {Shandarin}}]{Coles93}
{Coles}, P., {Melott}, A.~L., \& {Shandarin}, S.~F. 1993, \mnras, 260, 765

\bibitem[{{Crocce} {et~al.}(2006){Crocce}, {Pueblas}, \&
  {Scoccimarro}}]{Crocce06b}
{Crocce}, M., {Pueblas}, S., \& {Scoccimarro}, R. 2006, \mnras, 373, 369

\bibitem[{{Crocce} \& {Scoccimarro}(2006)}]{Crocce06a}
{Crocce}, M., \& {Scoccimarro}, R. 2006, \prd, 73, 063520

\bibitem[{{Efstathiou} {et~al.}(1985){Efstathiou}, {Davis}, {White}, \&
  {Frenk}}]{Efstathiou85}
{Efstathiou}, G., {Davis}, M., {White}, S.~D.~M., \& {Frenk}, C.~S. 1985,
  \apjs, 57, 241

\bibitem[{{Franx} {et~al.}(1991){Franx}, {Illingworth}, \& {de
  Zeeuw}}]{Franx91}
{Franx}, M., {Illingworth}, G., \& {de Zeeuw}, T. 1991, \apj, 383, 112

\bibitem[{{Frenk} {et~al.}(1999){Frenk}, {White}, {Bode}, {Bond}, {Bryan},
  {Cen}, {Couchman}, {Evrard}, {Gnedin}, {Jenkins}, {Khokhlov}, {Klypin},
  {Navarro}, {Norman}, {Ostriker}, {Owen}, {Pearce}, {Pen}, {Steinmetz},
  {Thomas}, {Villumsen}, {Wadsley}, {Warren}, {Xu}, \& {Yepes}}]{Frenk99}
{Frenk}, C.~S., {White}, S.~D.~M., {Bode}, P., {Bond}, J.~R., {Bryan}, G.~L.,
  {Cen}, R., {Couchman}, H.~M.~P., {Evrard}, A.~E., {Gnedin}, N., {Jenkins},
  A., {Khokhlov}, A.~M., {Klypin}, A., {Navarro}, J.~F., {Norman}, M.~L.,
  {Ostriker}, J.~P., {Owen}, J.~M., {Pearce}, F.~R., {Pen}, U.-L., {Steinmetz},
  M., {Thomas}, P.~A., {Villumsen}, J.~V., {Wadsley}, J.~W., {Warren}, M.~S.,
  {Xu}, G., \& {Yepes}, G. 1999, \apj, 525, 554

\bibitem[{{Gill} {et~al.}(2004){Gill}, {Knebe}, \& {Gibson}}]{Gill04a}
{Gill}, S.~P.~D., {Knebe}, A., \& {Gibson}, B.~K. 2004, \mnras, 351, 399

\bibitem[{{Hansen} {et~al.}(2007){Hansen}, {Agertz}, {Joyce}, {Stadel},
  {Moore}, \& {Potter}}]{Hansen07}
{Hansen}, S.~H., {Agertz}, O., {Joyce}, M., {Stadel}, J., {Moore}, B., \&
  {Potter}, D. 2007, \apj, 656, 631

\bibitem[{{Heitmann} {et~al.}(2007){Heitmann}, {Lukic}, {Fasel}, {Habib},
  {Warren}, {White}, {Ahrens}, {Ankeny}, {Armstrong}, {O'Shea}, {Ricker},
  {Springel}, {Stadel}, \& {Trac}}]{Heitmann07}
{Heitmann}, K., {Lukic}, Z., {Fasel}, P., {Habib}, S., {Warren}, M.~S.,
  {White}, M., {Ahrens}, J., {Ankeny}, L., {Armstrong}, R., {O'Shea}, B.,
  {Ricker}, P.~M., {Springel}, V., {Stadel}, J., \& {Trac}, H. 2007, ArXiv
  e-prints, 706

\bibitem[{{Heitmann} {et~al.}(2006){Heitmann}, {Luki{\'c}}, {Habib}, \&
  {Ricker}}]{Heitmann06}
{Heitmann}, K., {Luki{\'c}}, Z., {Habib}, S., \& {Ricker}, P.~M. 2006, \apjl,
  642, L85

\bibitem[{{Heitmann} {et~al.}(2008){Heitmann}, {White}, {Wagner}, {Habib}, \&
  {Higdon}}]{Heitmann08}
{Heitmann}, K., {White}, M., {Wagner}, C., {Habib}, S., \& {Higdon}, D. 2008,
  ArXiv e-prints

\bibitem[{{Hockney} \& {Eastwood}(1988)}]{Hockney88}
{Hockney}, R.~W., \& {Eastwood}, J.~W. 1988, {Computer simulation using
  particles} (Bristol: Hilger, 1988)

\bibitem[{{Jenkins} {et~al.}(2001){Jenkins}, {Frenk}, {White}, {Colberg},
  {Cole}, {Evrard}, {Couchman}, \& {Yoshida}}]{Jenkins01}
{Jenkins}, A., {Frenk}, C.~S., {White}, S.~D.~M., {Colberg}, J.~M., {Cole}, S.,
  {Evrard}, A.~E., {Couchman}, H.~M.~P., \& {Yoshida}, N. 2001, \mnras, 321,
  372

\bibitem[{{Joyce} \& {Marcos}(2007{\natexlab{a}})}]{Joyce07b}
{Joyce}, M., \& {Marcos}, B. 2007{\natexlab{a}}, \prd, 76, 103505

\bibitem[{{Joyce} \& {Marcos}(2007{\natexlab{b}})}]{Joyce07a}
---. 2007{\natexlab{b}}, \prd, 75, 063516

\bibitem[{{Joyce} {et~al.}(2008){Joyce}, {Marcos}, \& {Baertschiger}}]{Joyce08}
{Joyce}, M., {Marcos}, B., \& {Baertschiger}, T. 2008, ArXiv e-prints, 805

\bibitem[{{Klypin} \& {Shandarin}(1983)}]{Klypin83}
{Klypin}, A.~A., \& {Shandarin}, S.~F. 1983, \mnras, 204, 891

\bibitem[{{Knebe} \& {Dom{\'{\i}}nguez}(2003)}]{Knebe03}
{Knebe}, A., \& {Dom{\'{\i}}nguez}, A. 2003, Publications of the Astronomical
  Society of Australia, 20, 173

\bibitem[{{Knebe} {et~al.}(2006){Knebe}, {Dom{\'{\i}}nguez}, \&
  {Dom{\'{\i}}nguez-Tenreiro}}]{Knebe06happi}
{Knebe}, A., {Dom{\'{\i}}nguez}, A., \& {Dom{\'{\i}}nguez-Tenreiro}, R. 2006,
  \mnras, 371, 1959

\bibitem[{{Knebe} {et~al.}(2000){Knebe}, {Kravtsov}, {Gottl{\"o}ber}, \&
  {Klypin}}]{Knebe00}
{Knebe}, A., {Kravtsov}, A.~V., {Gottl{\"o}ber}, S., \& {Klypin}, A.~A. 2000,
  \mnras, 317, 630

\bibitem[{{Knollmann} \& {Knebe}(2009)}]{Knollmann09}
{Knollmann}, S., \& {Knebe}, A. 2009, \apj, in press, 0, 0

\bibitem[{{Komatsu} {et~al.}(2008){Komatsu}, {Dunkley}, {Nolta}, {Bennett},
  {Gold}, {Hinshaw}, {Jarosik}, {Larson}, {Limon}, {Page}, {Spergel},
  {Halpern}, {Hill}, {Kogut}, {Meyer}, {Tucker}, {Weiland}, {Wollack}, \&
  {Wright}}]{Komatsu08}
{Komatsu}, E., {Dunkley}, J., {Nolta}, M.~R., {Bennett}, C.~L., {Gold}, B.,
  {Hinshaw}, G., {Jarosik}, N., {Larson}, D., {Limon}, M., {Page}, L.,
  {Spergel}, D.~N., {Halpern}, M., {Hill}, R.~S., {Kogut}, A., {Meyer}, S.~S.,
  {Tucker}, G.~S., {Weiland}, J.~L., {Wollack}, E., \& {Wright}, E.~L. 2008,
  ArXiv e-prints, 803

\bibitem[{{Luki{\'c}} {et~al.}(2007){Luki{\'c}}, {Heitmann}, {Habib},
  {Bashinsky}, \& {Ricker}}]{Lukic07}
{Luki{\'c}}, Z., {Heitmann}, K., {Habib}, S., {Bashinsky}, S., \& {Ricker},
  P.~M. 2007, \apj, 671, 1160

\bibitem[{{Ma}(2007)}]{Ma07}
{Ma}, Z. 2007, \apj, 665, 887

\bibitem[{{Navarro} {et~al.}(1997){Navarro}, {Frenk}, \& {White}}]{Navarro97}
{Navarro}, J.~F., {Frenk}, C.~S., \& {White}, S.~D.~M. 1997, \apj, 490, 493

\bibitem[{{Nishimichi} {et~al.}(2008){Nishimichi}, {Shirata}, {Taruya},
  {Yahata}, {Saito}, {Suto}, {Takahashi}, {Yoshida}, {Matsubara}, {Sugiyama},
  {Kayo}, {Jing}, \& {Yoshikawa}}]{Nishimichi08}
{Nishimichi}, T., {Shirata}, A., {Taruya}, A., {Yahata}, K., {Saito}, S.,
  {Suto}, Y., {Takahashi}, R., {Yoshida}, N., {Matsubara}, T., {Sugiyama}, N.,
  {Kayo}, I., {Jing}, Y., \& {Yoshikawa}, K. 2008, ArXiv e-prints

\bibitem[{{O'Shea} {et~al.}(2005){O'Shea}, {Nagamine}, {Springel}, {Hernquist},
  \& {Norman}}]{OShea05}
{O'Shea}, B.~W., {Nagamine}, K., {Springel}, V., {Hernquist}, L., \& {Norman},
  M.~L. 2005, \apjs, 160, 1

\bibitem[{{Pen}(1997)}]{Pen97}
{Pen}, U.-L. 1997, \apjl, 490, L127+

\bibitem[{{Power} \& {Knebe}(2006)}]{Power06}
{Power}, C., \& {Knebe}, A. 2006, \mnras, 370, 691

\bibitem[{{Prunet} {et~al.}(2008){Prunet}, {Pichon}, {Aubert}, {Pogosyan},
  {Teyssier}, \& {Gottloeber}}]{Prunet08}
{Prunet}, S., {Pichon}, C., {Aubert}, D., {Pogosyan}, D., {Teyssier}, R., \&
  {Gottloeber}, S. 2008, \apjs, 178, 179

\bibitem[{{Reed} {et~al.}(2003){Reed}, {Gardner}, {Quinn}, {Stadel}, {Fardal},
  {Lake}, \& {Governato}}]{Reed03}
{Reed}, D., {Gardner}, J., {Quinn}, T., {Stadel}, J., {Fardal}, M., {Lake}, G.,
  \& {Governato}, F. 2003, \mnras, 346, 565

\bibitem[{{Reed} {et~al.}(2007){Reed}, {Bower}, {Frenk}, {Jenkins}, \&
  {Theuns}}]{Reed07}
{Reed}, D.~S., {Bower}, R., {Frenk}, C.~S., {Jenkins}, A., \& {Theuns}, T.
  2007, \mnras, 374, 2

\bibitem[{{Regan} {et~al.}(2007){Regan}, {Haehnelt}, \& {Viel}}]{Regan07}
{Regan}, J.~A., {Haehnelt}, M.~G., \& {Viel}, M. 2007, \mnras, 374, 196

\bibitem[{{Scoccimarro}(1998)}]{Scoccimarro98}
{Scoccimarro}, R. 1998, \mnras, 299, 1097

\bibitem[{{Sirko}(2005)}]{Sirko05}
{Sirko}, E. 2005, \apj, 634, 728

\bibitem[{{Splinter} {et~al.}(1998){Splinter}, {Melott}, {Shandarin}, \&
  {Suto}}]{Splinter98}
{Splinter}, R.~J., {Melott}, A.~L., {Shandarin}, S.~F., \& {Suto}, Y. 1998,
  \apj, 497, 38

\bibitem[{{Springel} {et~al.}(2008){Springel}, {Wang}, {Vogelsberger},
  {Ludlow}, {Jenkins}, {Helmi}, {Navarro}, {Frenk}, \& {White}}]{Springel08}
{Springel}, V., {Wang}, J., {Vogelsberger}, M., {Ludlow}, A., {Jenkins}, A.,
  {Helmi}, A., {Navarro}, J.~F., {Frenk}, C.~S., \& {White}, S.~D.~M. 2008,
  ArXiv e-prints, 809

\bibitem[{{Stadel} {et~al.}(2008){Stadel}, {Potter}, {Moore}, {Diemand},
  {Madau}, {Zemp}, {Kuhlen}, \& {Quilis}}]{Stadel08}
{Stadel}, J., {Potter}, D., {Moore}, B., {Diemand}, J., {Madau}, P., {Zemp},
  M., {Kuhlen}, M., \& {Quilis}, V. 2008, ArXiv e-prints, 808

\bibitem[{{Tasker} {et~al.}(2008){Tasker}, {Brunino}, {Mitchell}, {Michielsen},
  {Hopton}, {Pearce}, {Bryan}, \& {Theuns}}]{Tasker08}
{Tasker}, E.~J., {Brunino}, R., {Mitchell}, N.~L., {Michielsen}, D., {Hopton},
  S., {Pearce}, F.~R., {Bryan}, G.~L., \& {Theuns}, T. 2008, ArXiv e-prints,
  808

\bibitem[{{Tatekawa} \& {Mizuno}(2007)}]{Tatekawa07}
{Tatekawa}, T., \& {Mizuno}, S. 2007, Journal of Cosmology and Astro-Particle
  Physics, 12, 14

\bibitem[{{Tinker} {et~al.}(2008){Tinker}, {Kravtsov}, {Klypin}, {Abazajian},
  {Warren}, {Yepes}, {Gottl{\"o}ber}, \& {Holz}}]{Tinker08}
{Tinker}, J., {Kravtsov}, A.~V., {Klypin}, A., {Abazajian}, K., {Warren}, M.,
  {Yepes}, G., {Gottl{\"o}ber}, S., \& {Holz}, D.~E. 2008, \apj, 688, 709

\bibitem[{{Valageas}(2002)}]{Valageas02}
{Valageas}, P. 2002, \aap, 385, 761

\bibitem[{{Valluri} {et~al.}(2007){Valluri}, {Vass}, {Kazantzidis}, {Kravtsov},
  \& {Bohn}}]{Valluri07}
{Valluri}, M., {Vass}, I.~M., {Kazantzidis}, S., {Kravtsov}, A.~V., \& {Bohn},
  C.~L. 2007, \apj, 658, 731

\bibitem[{{Zel'dovich}(1970)}]{Zeldovich70}
{Zel'dovich}, Y.~B. 1970, \aap, 5, 84

\end{thebibliography}


\end{document}